\documentclass[twocolumn,english,prb,showpacs]{revtex4-1}
\usepackage[latin9]{inputenc}
\usepackage{amsmath}
\usepackage{amssymb}
\usepackage{graphicx}
\usepackage{esint}

\makeatletter
 
 \@ifundefined{textcolor}{}
 {%
   \definecolor{BLACK}{gray}{0}
   \definecolor{WHITE}{gray}{1}
   \definecolor{RED}{rgb}{1,0,0}
   \definecolor{GREEN}{rgb}{0,1,0}
   \definecolor{BLUE}{rgb}{0,0,1}
   \definecolor{CYAN}{cmyk}{1,0,0,0}
   \definecolor{MAGENTA}{cmyk}{0,1,0,0}
   \definecolor{YELLOW}{cmyk}{0,0,1,0}
 }

\newcommand{\Imm}{\mathrm{Im}}
\newcommand{\Ree}{\mathrm{Re}}

\usepackage{babel}

\usepackage{babel}

\makeatother

\usepackage{babel}
\begin{document}

\title{Plasmon anomaly in the dynamical optical conductivity of graphene}

\author{K. Kechedzhi and S. Das Sarma}

\affiliation{Condensed Matter Theory Center and Joint Quantum Institute, Department
of Physics, University of Maryland College Park, Maryland 20742-4111,
USA}
\begin{abstract}
We theoretically consider the effect of plasmon collective modes on
the frequency-dependent conductivity of graphene in the presence of
the random static potential of charged impurities. We develop an equation
of motion approach suitable for the relativistic Dirac electrons in
graphene that allows analytical high-frequency asymptotic solution
($\omega\tau\gg1$ where $\tau$ is the scattering time) in the presence
of both disorder and interaction. We show that the presence of the
gapless plasmon pole (in graphene the plasmon frequency vanishes at
long wavelengths as the square-root of wavenumber) in the inverse
dynamical dielectric function of graphene gives rise to a strong variation
with frequency of the screening effect of the relativistic electron
gas in graphene on the potential of charged impurities. The resulting
frequency-dependent impurity scattering rate gives rise to a broad
peak in the frequency-dependent graphene optical conductivity with
the amplitude and the position of the peak being sensitive to the
detailed characteristics of disorder and interaction in the system.
This sample-dependent (i.e., disorder, electron density, and interaction
strength) redistribution of the spectral weight in the frequency-dependent
graphene conductivity may have already been experimentally observed
in optical measurements. 
\end{abstract}
\maketitle

\section{introduction}

Electronic transport and electrical conductivity of graphene is a
subject \cite{das_sarma_electronic_2011} of great current interest
in both fundamental physics and in applied physics and engineering.
Recently, optical conductivity measurements have been used to identify
the effects of electron-electron interactions of relativistic Dirac
electrons\ \cite{horng_drude_2011,mak_optical_2012} which have attracted
a great deal of interest. In particular, the linear band dispersion
as well as its gapless chiral 2D nature make graphene a unique laboratory
system for studying electron-electron interaction effects and their
interplay with disorder since the lack of Galilean invariance associated
with the non-classical linear band dispersion leads to qualitative
and quantitative phenomena not present in regular parabolic band metals
and semiconductors. For example, graphene violates the well-known
Kohn's theorem, and both cyclotron resonance and optical conductivity
are directly affected by electron-electron interaction effects in
contrast to the corresponding parabolic band systems. In the context
of engineering applications, strong broadband absorption and high
tunability make monolayer graphene uniquely suited for a wide range
of optoelectronic and plasmonic devices\ \cite{grigorenko_graphene_2012}.
These prospects have motivated numerous measurements of the optical
conductivity of monolayer graphene in the wide frequency range from
the far infrared to the ultraviolet\ \cite{frenzel_observation_2013,ren_terahertz_2012,maeng_gate-controlled_2012,george_ultrafast_2008,choi_broadband_2009,jiang_quantum_2010,mak_measurement_2008,yan_infrared_2011}.
A detailed understanding of the optical properties of monolayer graphene
is required for the progress in both of these directions. In this
paper we consider only monolayer graphene within the effective linear
gapless Dirac band model, and 'graphene' in this article refers exclusively
to 'monolayer graphene'. Our goal here is to theoretically study the
effects of electronic collective modes on the graphene dynamical conductivity
treating disorder and interaction on the same footing.

\subsection{Experimental motivation}

The optical conductivity of undoped graphene is dominated by interband
processes (arising from the particle-hole excitations from the filled
valence band to the empty conduction band) in the gapless Dirac spectrum.
In the non-interacting case this gives rise to a universal frequency-independent
value of the conductivity that can be expressed in terms of fundamental
constants, a prediction\ \cite{Ando02OptCond,gusynin_unusual_2006}
remarkably confirmed by the measurements\ \cite{nair_fine_2008,mak_measurement_2008},
establishing the basic gapless chiral massless Dirac band dispersion
of graphene to be valid. At finite doping the interband processes
are strongly suppressed by Pauli blocking in the partially filled
conduction band (assuming positive Fermi level $E_{F}>0$, but obviously
the same physics also applies for the hole-doped system) whereas the
universal value of the conductivity is reached at higher frequencies
$\hbar\omega\gtrsim2E_{F}$ where Pauli blocking effects become small,
and the interband transitions again dominate the optical conductivity.
The partially filled conduction band gives rise to the intraband response
reflected in the Drude peak at $\omega\approx0$ in the optical conductivity.
In graphene electron-electron interactions are expected to give rise
to relatively small yet observable corrections to the optical conductivity
via the many-body renormalization of the electron spectrum and the
excitonic effects near the absorption threshold $\hbar\omega=2E_{F}$\ \cite{das_sarma_electronic_2011,mak_optical_2012}.
An important difference between graphene and ordinary parabolic band
semiconductor systems in this context is that the linearly dispersing
graphene electron and hole bands violate the usual Galilean invariance
dominating the long wavelength optical response of the semiconductor
systems where the electron-electron interaction, being a property
of the relative coordinates of the individual electrons themselves,
does not affect the long wavelength translationally invariant optical
conductivity of the system since the latter is a specific property
of the center of mass coordinates coupling to the external light in
the long wavelength limit. The long wavelength optical properties
in graphene, however, are explicitly affected by the interaction effects
since graphene obeys Lorentz invariance which does not enable the
usual separation of the system Hamiltonian into the center of mass
and relative coordinates components.

In particular, the Drude weight in graphene has been predicted to
be renormalized by the interaction effects\ \cite{McDonaldEnhancedDrudeWeight}.
Experimentally this effect is demonstrated by analyzing the parameters
of the Drude peak. The latter is routinely phenomenologically fitted
by a 2-parameter Lorentzian $\sigma=iD/(\pi(\omega+i/\tau))$ characterized
by a width $1/\tau$ due to the disorder scattering and the Drude
weight $D$\ \cite{horng_drude_2011,mak_optical_2012}. The latter
has shown discrepancy with the free-electron value $D=(ve^{2}/\hbar)\sqrt{\pi\rho}$
suggestive of electron-electron interactions playing a role. Here
$v$ is the slope of the linear dispersion in graphene (i.e. the so
called graphene velocity) and $\rho$ is the electron density. On
the other hand, the data reported in Ref.~\onlinecite{BasovFIR08}
suggest a non-Lorentzian shape of the Drude peak in graphene possibly
reflecting a frequency-dependent relaxation rate $1/\tau(\omega)$
or the presence of other non-universal background at $\omega\tau\gtrsim1$.
More recently, additional measurements of the plasmon decay\ \cite{Nanoscopy11,Basov12},
directly related to the optical conductivity, confirmed the discrepancies
with the simple Lorentzian fits. The questions of the origin of the
non-Lorentzian shape of the Drude peak and the actual value of the
Drude weight extracted from the Drude response remain open.

\subsection{Theoretical motivation}

Several theoretical models of the optical response of graphene have
been put forward that include the effects of disorder\ \cite{Ando02OptCond,EffectOfVariousDisorderOnConduct,ashby_tracking_2012},
the band structure corrections due to the next-nearest neighbor hopping
in the tight binding description\ \cite{ConductAtHighDensity}, excitonic
effects near $\hbar\omega\approx2E_{F}$\ \cite{ExcitonicEffectsOnConduct,CastroNetoOptCond10},
effects of phonons\ \cite{PhononEffectOnConduct,EffectOfPhononOnConduct,DisorderAndPhononEffectOnConduct,ElectronElectronInteractionGapPhononsOnConduct,Carbotte12,carbotte_impact_2013}
relevant at high temperatures, plasmaron effects\ \cite{hwang_quasiparticle_2008,*CarbottePalsmaron}
near the Dirac point, the effect of strain\ \cite{StrainEffectOnConduct}
and the spectrum renormalization effects of electron-electron interactions\ \cite{Carbotte12,Vozmediano09}.

An effect omitted so far in the literature is the frequency dependence
of the Lorentz invariant screening effect of the graphene relativistic
electron gas. Formally, the dynamical screening of an electron gas
is characterized by the inverse of the dynamical dielectric function,
$1/\varepsilon(q,\omega)$. The presence of a plasmon pole in the
dynamical dielectric function, $\Imm m[1/\varepsilon(q,\omega)]\sim\delta(\omega-\omega_{pl}(q))$,
results in the failure of the screening at the plasmon frequency $\omega\approx\omega_{pl}(q)$
and therefore gives rise to an enhancement of the disorder scattering
rate. As a result, the gapless plasmon dispersion $\omega_{pl}(q)$
(i.e. the plasma frequency going as the square-root of the wavenumber
at long wavelength) in the 2D electron gas (2DEG) gives rise to a
strong frequency dependence of the scattering rate due to charged
impurities\ \cite{GotzeWolfle72,BelitzDasSarma,gold_localization_1986,gold_enhanced_1990}.

Despite the conceptual simplicity, a formal calculation of the frequency-dependent
scattering rate in the electron gas in the presence of disorder and
interaction is not straightforward. The standard method is to construct
a diagrammatic perturbation theory for the current-current correlator
relying on the ``metallic'' expansion parameter $E_{F}\tau\gg\hbar$
characterizing the disorder strength and then employing some sort
of approximation scheme for the electron-electron interactions. Such
a construction however requires the inclusion of an infinite series
of diagrams. The summation of the infinite series results in an integral
Bethe-Salpeter equation for the vertex function which is difficult
and cumbersome to solve in the presence of both disorder and interaction
even for the simplest approximation schemes.

In the case of the 2DEG with the parabolic dispersion the problem
was successfully solved with a lot of work using the so-called 'memory
function' approach\ \cite{GotzeWolfle72}. This solution relied on
a substantial simplification that is possible in the parabolic case
due to the separation between the slow dynamics of the center of mass
of the electron gas and the relative motion of electrons. As a result
the dynamics of the 2DEG is described by an effective quantum Langevin
equation for a macroscopic particle with a large mass (the mass of
the electron gas) subject to a random force due to the disorder potential\ \cite{hu_memory_1988}.
In this case the force-force correlator determines the response of
the 2DEG to the external field. Electron-electron interactions enter
the force-force correlator in a particularly simple way, only via
the polarization operator, which greatly simplifies the calculation\ \cite{GotzeWolfle72}.

In graphene the situation is more challenging. Due to the broken Galilean
invariance the dynamics of the center of mass of the electron gas
is coupled to the pseudospin degree of freedom\ \cite{muller_collective_2008,McDonaldEnhancedDrudeWeight}.
This is reflected in the violation of Kohn's theorem in graphene.
Therefore one has to take great care when applying the quantum Langevin
equation approach to graphene. In particular, in the presence of electron-electron
interactions the force-force correlator does not have the simple form
of the parabolic case. Therefore the Langevin equation does not readily
provide any advantage over the standard Bethe-Salpeter equation.

The situation is greatly simplified in the high frequency regime $\omega\tau\gg1$
where the summation of the infinite series of diagrams in powers of
the disorder potential is not required. In this paper we adopt for
the case of the Dirac electrons in graphene an equation of motion
framework, developed in Refs.~\onlinecite{LuttingerKohn,tzoarhigh-frequency1985},
which is suitable for the high frequency regime. This approach is
somewhat similar to the derivation of the quantum kinetic equation\ \cite{auslender_generalized_2007,fritz_quantum_2008,muller_collective_2008,kechedzhi_quantum_2008},
however, here we will employ an equal time correlator, the 'density
matrix', as opposed to the quasiclassical Green's function used in
the former approach.

\subsection{Summary of the calculation and results }

In this paper we develop an equation of motion approach in order to
calculate the high frequency asymptotic, $1/(\omega\tau)\ll1$, (accessible
in current experiments%
\footnote{For comparison, parameters of the plasmon decay measurements in Ref.~\onlinecite{Basov12}
correspond to $\omega\tau\sim5$ so that the regime considered in
this paper is accessible in current experiments.%
}) of the optical conductivity of graphene in the presence of electron-electron
interactions and charged impurities. Interaction effects are included
in the self-consistent mean field approximation which is equivalent
to the random phase approximation (RPA) of the diagrammatic perturbation
theory.

We find an enhancement of the disorder scattering rate at finite frequencies
due to the presence of the graphene plasmon excitation in the spectrum.
We make a quantitative prediction for the frequency-dependent correction
to the optical conductivity of graphene due to the plasmon anomaly
in the perturbative regime, $\hbar/\tau\ll\hbar\omega\ll E_{F}$.
In the following we use two frequency-dependent expansion parameters,
$\omega\tau\gg1$ and $\omega/E_{F}\ll1$. The former will be referred
to as the high frequency asymptotic and the regime $\omega\tau\lesssim1$
will be referred to as low frequency regime. We also discuss the extension
of our results beyond this frequency range. The plasmon enhanced scattering
gives rise to a broad peak in the frequency-dependent conductivity
given by, 
\begin{eqnarray*}
 & \Ree\sigma(\omega)\propto\left(\frac{\hbar\omega}{cE_{F}}\right)^{3}\exp\left[-\left(\frac{\hbar\omega}{cE_{F}}\right)^{2}\right].
\end{eqnarray*}
The dimensionless coefficient $c\equiv\sqrt{\alpha/(k_{F}d)}$, depends
on the Fermi wave vector $k_{F}$ and for typical values of parameters
can be small $c\ll1$. Here $d$ is the effective separation of the
charged impurities from the graphene plane, and $\alpha\equiv e^{2}/(\kappa v\hbar)$
is the dimensionless interaction strength (i.e. the so called graphene
fine structure constant) and $\kappa$ is the effective dielectric
constant. The rapid increase of the scattering rate at $\hbar\omega\gtrsim cE_{F}$
is limited by the flattening of the Coulomb potential of the impurities
at the length scale $d$. Both the location $\hbar\omega^{\ast}/E_{F}=\sqrt{3/2}c$
and the magnitude $\Ree\sigma(\omega^{\ast})$ of the peak are sensitive
to the characteristics of disorder and the Fermi level $E_{F}$ in
the system. Therefore the effect is expected to be strongly sample
dependent (i.e. impurity distribution, carrier density and interaction
strength).

We use our perturbative high frequency results to make a qualitative
prediction for the frequency-dependent optical conductivity at low
frequencies $\omega\tau\lesssim1$. We find that in this regime the
plasmon induced frequency-dependent scattering rate leads to a redistribution
of the spectral weight in the optical conductivity, which may be substantial
in the low density regime in strongly disordered samples. The redistribution
of the spectral weight results in the non-Lorentzian shape of the
Drude peak which has to be accounted for when extracting various system
parameters from measurements of the Drude response.

The paper is organized as follows. In Sec.\ \ref{sec:setup} we discuss
the low energy Hamiltonian of monolayer graphene and the model of
Coulomb disorder arising from charged defects. An equation of motion
is derived in Sec.\ \ref{sec:Equation-of-Motion} and solved in Sec.\ \ref{sec:EOMSolution}
and details given in Appendix\ \ref{sub:Full-solution}. The plasmon
propagator is introduced in Sec.\ \ref{sec:Plasmon-propagator}.
Asymptotic solutions of the equation of motion presented in Sec.\ \ref{sec:Asymptotic-solutions}
are used to obtain quantitative predictions for the optical conductivity
in Sec.\ \ref{sec:Results}. We conclude in Sec.\ \ref{sec:Discussion}.
Appendix\ \ref{sec:Appendix-I} contains a description of the formal
theory of the low-energy excitations in graphene that includes a band
cut off and Appendix\ \ref{sub:Full-solution} gives some details
of the derivation for the results in Secs.\ \ref{sec:EOMSolution}
and \ref{sec:Asymptotic-solutions}.

\section{model setup\label{sec:setup}}

The unit cell of the hexagonal lattice of monolayer graphene contains
two chemically equivalent carbon atoms (labeled $A$ and $B$). The
resulting degeneracy ensures the crossing of the conduction and valence
bands at the corners of the hexagonal Brillouin zone (labeled $K$
and $K'$). The low energy excitations ($<1\textrm{eV}$) are described
by the two flavor Dirac Hamiltonian, 
\begin{eqnarray*}
 & H_{0}=\sum_{\mathbf{k}}\hat{\Psi}_{\mathbf{k}}^{\dagger}\hat{h}_{\mathbf{k}}\hat{\Psi}_{\mathbf{k}},\\
 & \hat{h}_{\mathbf{k}}=v\Sigma\cdot\mathbf{k},
\end{eqnarray*}
with $\Sigma^{x/y/z}$ being the $4\times4$ matrices in the sublattice
and valley space that form a Pauli matrix algebra $\left[\Sigma^{j},\Sigma^{l}\right]=2i\mathcal{E}_{jlm}\Sigma^{m}$
where $\mathcal{E}{}_{jlm}$ is the antisymmetric tensor. The basis
chosen here is $\hat{\Psi}_{\mathbf{k}}^{\dagger}=[\psi_{AK\eta}^{\dagger}\psi_{BK\eta}^{\dagger}\psi_{BK'\eta}^{\dagger}\psi_{AK'\eta}^{\dagger}]$
where $\psi_{AK\eta}^{\dagger}$ and $\psi_{AK\eta}$ are the creation
and annihilation operators for an electron on the sublattice $A$
at the $K$ point, characterized by momentum $\mathbf{k}$ and spin
$\eta$. Below we assume that the Hamiltonian of graphene is diagonal
in spin indexes (and a spin degeneracy of 2 as well as a valley degeneracy
of 2 will be assumed throughout). Throughout the text we set $\hbar\equiv1$
and restore it in the final answers presented in Sec.\ \ref{sec:Results},
unless explicitly stated. The full low-energy Hamiltonian,

\begin{equation}
H=H_{0}+H_{ee}+H_{imp},\label{eq:H}
\end{equation}
includes the inter-electron Coulomb interaction, 
\begin{equation}
H_{ee}=\frac{1}{2S}\sum_{\mathbf{k,k',q}\neq0}V_{q}\Psi_{\beta\mathbf{k}}^{\dagger}\Psi_{\gamma\mathbf{k'}}^{\dagger}\Psi_{\gamma\mathbf{k'+q}}\Psi_{\beta\mathbf{k-q}},
\end{equation}
with $V_{q}\equiv2\pi e^{2}/(\kappa q)$, characterized by a dimensionless
ratio of the typical Coulomb energy to the typical kinetic energy
$\alpha=\frac{e^{2}}{\kappa v}$ (here $\hbar\equiv1$) that is independent
of the electron density. The full Hamiltonian also includes the potential
of random charged impurities, 
\begin{eqnarray}
 & H_{imp}=\sum_{\mathbf{k,q}}V_{\mathbf{q}}^{(i)}\Psi_{\beta\mathbf{k}}^{\dagger}\Psi_{\beta\mathbf{k+q}},\\
 & V_{\mathbf{q}}^{(i)}=\sum_{i}V_{q}e^{i\mathbf{q}\mathbf{r_{i}}-qd_{i}},
\end{eqnarray}
which dominates the elastic disorder scattering in typical graphene
devices fabricated on $\textrm{Si\ensuremath{O_{2}}/Si}$ substrate.
Here $\mathbf{q}$ is a 2D momentum in the graphene plane, $r_{i}$
is the location of the $i$th impurity (assumed random) in the 2D
plane parallel to the graphene layer, and $d_{i}$ is the distance
between the graphene layer and the impurity in the direction perpendicular
to the graphene plane. For simplicity, and with no loss of generality
we assume that all impurities are located in a 2D plane parallel to
the graphene layer located a distance $d$ away from it. We will comment
on the implications of the impurities being distributed in three dimensions
later in the paper. We assume without loss of generality the distribution
of impurities to be charge-neutral such that the average of the disorder
potential over the random disorder configurations vanishes $\overline{\left(V^{(i)}(r)\right)}=0$.
We use the Born approximation for the disorder scattering rate such
that the disorder scattering is fully characterized by the correlator
of the disorder potential averaged over the random disorder configurations,

\begin{equation}
\overline{\left(V_{\mathbf{q}}^{(i)}V_{\mathbf{q'}}^{(i)}\right)}=\delta_{\mathbf{q,-q'}}\rho_{imp}V_{q}^{2}e^{-2qd},\label{VCorr}
\end{equation}
where $\delta_{\mathbf{q,q'}}$ stands for Kronecker symbol. Our model of uncorrelated
random disorder can be straightforwardly generalized to correlated
disorder scattering\ \cite{li_theory_2011} if experimental information
about disorder correlations is available. To keep the number of parameters
a minimum, we assume the impurity-induced random disorder to be completely
characterized by just two parameters, the impurity density $\rho_{imp}$
and their location with respect to the graphene layer $d$, which
is the minimal possible model for Coulomb disorder.

\section{Equation of Motion\label{sec:Equation-of-Motion}}

We introduce an equal time correlator 
\[
\hat{g}_{\mathbf{k,k+q}}(t)\equiv\langle\hat{\Psi}_{\mathbf{k}}^{\dagger}(t)\hat{\Psi}_{\mathbf{k+q}}(t)\rangle
\]
which will be called the 'density matrix' in this paper. The density
matrix, $\hat{g}_{\mathbf{k,k+q}}(t),$ is a $8\times8$ matrix defined
in terms of the spinor creation/annihilation operators $\hat{\Psi}_{\mathbf{k}}^{\dagger}$
in the basis of the Bloch wave functions labeled by the sublattice,
valley and spin indices. The density matrix is diagonal in the spin
indices which will give rise to a factor of $\eta=2$ each time the
trace is taken.

In the Heisenberg representation a time derivative of an operator
$-i\partial_{t}\hat{g}_{\mathbf{k,k+q}}(t)$ is given by a commutator
with the full Hamiltonian,

\begin{equation}
-i\partial_{t}\hat{g}_{\mathbf{k,k+q}}(t)=\left[H,\hat{g}_{\mathbf{k,k+q}}(t)\right].
\end{equation}
Calculating the commutators of the density matrix with the Hamiltonian
in Eq.\ (\ref{eq:H}) (see also Appendix\ \ref{sec:Appendix-I})
we find,

\begin{eqnarray}
 & \left[H_{0},\hat{g}_{\mathbf{k,k+q}}(t)\right]=\hat{h}_{\mathbf{k}}\hat{g}_{\mathbf{k,k+q}}(t)-\hat{g}_{\mathbf{k,k+q}}(t)\hat{h}_{\mathbf{k+q}},\label{eq:[g,H0]}
\end{eqnarray}
and,

\begin{eqnarray}
 & \left[H_{imp},\hat{g}_{\mathbf{k,k+q}}(t)\right]\nonumber \\
 & =\sum_{\mathbf{q'}}V_{\mathbf{q'}}^{(i)}\left(\hat{g}_{\mathbf{k-q',k+q}}(t)-\hat{g}_{\mathbf{k,k+q+q'}}(t)\right).\label{eq:[g,Himp]}
\end{eqnarray}
The interactions are included in the mean field approximation,

\begin{eqnarray}
 & \left[H_{ee},\hat{g}_{\mathbf{k,k+q}}(t)\right]\nonumber \\
 & =\sum_{\mathbf{q'}}V_{q'}\rho(-\mathbf{q'})\left(\hat{g}_{\mathbf{k-q',k+q}}(t)-\hat{g}_{\mathbf{k,k+q+q}'}(t)\right),\label{eq:[g,Hee]}
\end{eqnarray}
where $\rho(q)\equiv\sum_{k}\mathrm{Tr}\hat{g}(\mathbf{k},\mathbf{k+q},t)$
is the Fourier transform of the spatial fluctuation of the electron
density.

We include the external electric field in a gauge invariant formulation\ \cite{KineticBook},
which requires the monochromatic field $\mathbf{E}(t)=\mathbf{E}e^{i\omega t}$
to enter as follows,

\begin{equation}
\left(-\omega-ie\mathbf{E}\cdot\nabla_{\mathbf{k}}\right)\hat{g}_{\mathbf{k,k+q}}(\omega)=\left[H,\hat{g}_{\mathbf{k,k+q}}(\omega)\right],\label{eq:EOMomega}
\end{equation}
where we also take the Fourier transform with respect to time. Here
and throughout the text we assume that the frequency $\omega>0$ has
an infinitesimal positive imaginary part which is set to zero at the
end in the usual manner.

\subsection{Diagonalized equation of motion}

We introduce a projection operator,

\begin{equation}
\mathcal{P}_{s\mathbf{n_{k}}}\equiv\frac{1}{2}\left(1+s\Sigma\cdot\mathbf{n_{k}}\right),\label{eq:Projector}
\end{equation}
that projects on the subspace of positive and negative energy eigenstates
of the Dirac Hamiltonian $H_{0}$ for $s=1$ and $s=-1$, respectively.
Here $\mathbf{n_{k}\equiv\mathbf{k}/}k$. It can be verified directly
that $\mathcal{P}_{s\mathbf{n_{k}}}^{2}=\mathcal{P}_{s\mathbf{n_{k}}}$
and $\mathcal{P}_{\mathbf{n_{k}}}+\mathcal{P}_{-\mathbf{n_{k}}}=1$.

Multiplying both sides of Eq.\ (\ref{eq:EOMomega}) by $\mathcal{P}_{s\mathbf{n_{k}}}$
on the left and by $\mathcal{P}_{s'\mathbf{n_{k+q}}}$on the right
we diagonalize the free particle part Eq.\ (\ref{eq:[g,H0]}) in
the right hand side of the equation of motion Eq.\ (\ref{eq:EOMomega})
,

\begin{eqnarray*}
 & \mathcal{P}_{s\mathbf{n_{k}}}\left(\hat{h}_{\mathbf{k}}\hat{g}_{\mathbf{k,k+q}}(\omega)-\hat{g}_{\mathbf{k,k+q}}(\omega)\hat{h}_{\mathbf{k+q}}\right)\mathcal{P}_{s'\mathbf{n_{k+q}}}\\
 & =\left(s\epsilon_{\mathbf{k}}-s'\epsilon_{\mathbf{k+q}}\right)\mathcal{P}_{s\mathbf{n_{k}}}\hat{g}_{\mathbf{k,k+q}}(\omega)\mathcal{P}_{s'\mathbf{n_{k+q}}}.
\end{eqnarray*}
where we used that $\mathcal{P}_{s\mathbf{n_{k}}}v\Sigma\cdot\mathbf{k}=s\epsilon_{\mathbf{k}}\mathcal{P}_{s\mathbf{n_{k}}}$
and $\epsilon_{\mathbf{k}}\equiv vk$. We use an identity, 
\[
\hat{g}_{\mathbf{k,k+q}}(\omega)=\sum_{s,s'}\mathcal{P}_{s\mathbf{n_{k}}}\hat{g}_{\mathbf{k,k+q}}(\omega)\mathcal{P}_{s'\mathbf{n_{k+q}}},
\]
which can be verified directly, in the left hand side of Eq.\ (\ref{eq:EOMomega})
multiplied by the projectors (as described above) to obtain,

\begin{eqnarray}
 & \hat{g}_{\mathbf{k,k+q}}(\omega)=\sum_{ss'}\frac{\mathcal{P}_{s\mathbf{n_{k}}}\left[H_{ee}+H_{imp},\hat{g}_{\mathbf{k,k+q}}(\omega)\right]\mathcal{P}_{s'\mathbf{n_{k+q}}}}{s'\epsilon_{\mathbf{k+q}}-s\epsilon_{\mathbf{k}}-\omega}\nonumber \\
 & +\sum_{ss'}\frac{\mathcal{P}_{s\mathbf{n_{k}}}\left(ie\mathbf{E}\cdot\nabla_{\mathbf{k}}\hat{g}_{\mathbf{k,k+q}}(\omega)\right)\mathcal{P}_{s'\mathbf{n_{k+q}}}}{s'\epsilon_{\mathbf{k+q}}-s\epsilon_{\mathbf{k}}-\omega},\label{eq:EOMomegaDiag}
\end{eqnarray}
where all the terms containing perturbations are collected in the
right hand side which will be more convenient for the construction
of the perturbation theory.

\section{Solution of the equation of motion\label{sec:EOMSolution}}

The solution of the matrix integro-differential equation\ (\ref{eq:EOMomegaDiag})
can be used to obtain the current response,

\begin{equation}
J_{total}^{j}=ev\sum_{k}\textrm{Tr}\{\Sigma^{j}\hat{g}_{\mathbf{k,k}}(\omega)\},\label{eq:CurrentGeneral}
\end{equation}
where $j=x,y$. Here and throughout the text we take the trace over
the sublattice, valley and spin indexes. In the linear response $J_{total}^{i}=\sigma_{ij}E_{j}$,
$i,j=x,y,$ and therefore it will be sufficient to keep the electric
field to the first order in the solution of\ (\ref{eq:EOMomegaDiag}).
The conductivity averaged over random disorder configurations is isotropic,
nevertheless, it will be more convenient for us to discuss components
of the current response and gives the results for the isotropic conductivity
at the end.

Solving the equation\ (\ref{eq:EOMomegaDiag}) is still a daunting
task as both the disorder potential and interactions have to be included
self-consistently. In the present work we solve this equation only
in the high-frequency regime $\omega\tau\gg1$ where $\tau$ is the
disorder scattering time. In this high-frequency regime it is sufficient
to keep the terms up to the second order in the disorder potential.
We treat electron-electron interactions in the self-consistent mean-field
approximation. This approach is equivalent to the standard random
phase approximation (RPA) of the many-body perturbation theory which
has been shown to provide an accurate description of the plasmon dispersion
in graphene\ \cite{das_sarma_collective_2009}. We neglect self-energy
and vertex corrections due to Coulomb interactions in the plasmon
dispersion because of the relatively weak electron-electron interaction
strength in graphene. We emphasize that the long-wavelength square
root in wave number plasmon dispersion in graphene is protected by
current conservation, but the coefficient of this long-wavelength
dispersion term is affected (weakly) by interaction effects in graphene
(in contrast to parabolic band systems), because of the violation
of Galilean invariance in graphene\ \cite{levitov_electron-electron_2013}.

We introduce a notation for the expansion of the density matrix,

\begin{eqnarray}
 & \hat{g}=\hat{g}^{(0)}+\hat{g}^{(10)}+\hat{g}^{(01)}+\hat{g}^{(02)}+\hat{g}^{(11)},\label{eq:g-expansion}
\end{eqnarray}
where $\hat{g}^{(ij)}$ corresponds to the $i$th-order expansion
in the electric field and the $j$th-order in the disorder potential,
and we drop the subscripts of $\hat{g}_{\mathbf{k,k+q}}^{(ij)}(\omega)$
in Eq.\ (\ref{eq:g-expansion}). Note that due to the violation of
the Kohn's theorem by the Dirac electrons, the external electric field
couples to the homogeneous electron density, $\hat{g}_{\mathbf{k,k+q}}^{(10)}(\omega)\neq0$,
in contrast to the ordinary parabolic band spectrum 2D electron systems\ \cite{tzoar_high-frequency_1985}.

In the following we find successive approximations of $\hat{g}_{\mathbf{k,k+q}}(\omega)$
defined in\ (\ref{eq:g-expansion}) by recursively solving Eq.\ (\ref{eq:EOMomegaDiag})
in each order of the perturbation theory. In the zeroth order, free
Dirac quasiparticles are described by the Fermi-Dirac distribution
$f_{s\epsilon_{\mathbf{k}}}=\frac{1}{1+e^{\beta s\epsilon_{\mathbf{k}}}}$,
and the density matrix is proportional to the projector Eq.\ (\ref{eq:Projector}),

\[
\hat{g}_{\mathbf{k,k+q}}^{(0)}\equiv\hat{g}_{\mathbf{k}}^{(0)}=\delta_{\mathbf{k,k+q}}\sum_{s=\pm}f_{s\epsilon_{\mathbf{k}}}\mathcal{P}_{s\mathbf{n_{k}}},
\]
where $\beta=1/(k_{B}T)$ is the inverse temperature and $k_{B}$
is the Boltzmann constant, and $\delta_{\mathbf{k,k+q}}$ stands for
the Kronecker symbol.

\subsection{First order terms}

The linear response of disorder free graphene to an external electric
field is described by $\hat{g}_{\mathbf{k,k+q}}^{(10)}(\omega),$
which satisfies,

\begin{eqnarray}
 & \hat{g}_{\mathbf{k,k+q}}(\omega)=\sum_{ss'}\frac{\mathcal{P}_{s\mathbf{n_{k}}}\left[H_{ee},\hat{g}_{\mathbf{k,k+q}}(\omega)\right]\mathcal{P}_{s'\mathbf{n_{k+q}}}}{s'\epsilon_{\mathbf{k+q}}-s\epsilon_{\mathbf{k}}-\omega}\nonumber \\
 & +\sum_{ss'}\frac{\mathcal{P}_{s\mathbf{n_{k}}}\left(ie\mathbf{E}\cdot\nabla_{\mathbf{k}}\hat{g}_{\mathbf{k,k+q}}(\omega)\right)\mathcal{P}_{s'\mathbf{n_{k+q}}}}{s'\epsilon_{\mathbf{k+q}}-s\epsilon_{\mathbf{k}}-\omega}.\label{eq:EOMFree}
\end{eqnarray}
We first solve Eq.\ (\ref{eq:EOMFree}) ignoring the Coulomb interaction.
In this case, in the first order in the electric field we obtain,

\begin{eqnarray}
 & \hat{g}_{\mathbf{k,k+q}}^{(10)}(\omega)=ie\mathbf{E}_{0}\cdot\sum\frac{\mathcal{P}_{s\mathbf{n_{k}}}\left(\nabla_{\mathbf{k}}\hat{g}_{\mathbf{k}}^{(0)}\right)\mathcal{P}_{s'\mathbf{n_{k}}}}{-\omega+(s'-s)\epsilon_{\mathbf{k}}}\delta_{\mathbf{k,k+q}},\label{eq:g10}
\end{eqnarray}
where the sum is over $s,s'=\pm1$. From this we directly obtain that
the electron density response to the homogeneous electric field, 
\begin{eqnarray*}
 & \rho^{(10)}(q,\omega)\equiv\sum_{\mathbf{k}}\mathrm{Tr}\hat{g}_{\mathbf{k,k+q}}^{(10)}(\omega)=0,
\end{eqnarray*}
vanishes in the linear order. Using Eqs.\ (\ref{eq:g10}) and\ (\ref{eq:CurrentGeneral})
we reproduce the universal value\ \cite{Ando02OptCond,gusynin_unusual_2006}
of the conductivity of non-interacting disorder free graphene at the
charge neutrality point ($E_{F}=0$),

\begin{eqnarray*}
 & \Ree\sigma_{E_{F}=0}=\frac{e^{2}}{4\hbar},
\end{eqnarray*}
where we restore the Planck's constant. At finite doping Eq.\ (\ref{eq:g10})
describes the optical absorption due to interband transitions with
energies $\omega>2E_{F}$ as well as the disorder-free zero-frequency
Drude peak\ \cite{McDonaldEnhancedDrudeWeight}, 
\begin{eqnarray*}
 & \Ree\sigma=\frac{e^{2}}{\hbar}E_{F}\delta(\omega),
\end{eqnarray*}
where we restore Planck's constant.

In general, electron-electron interactions give rise to corrections
to Eq.\ (\ref{eq:g10}) reflected in the many-body renormalization
of the Drude weight\ \cite{McDonaldEnhancedDrudeWeight} and the
renormalization of the universal conductivity $\sigma_{E_{F}=0}$\ \cite{sodemann_interaction_2012}.
Accounting for these renormalization effects would require including
self-energies and vertex corrections in the expansion in the electron-electron
interaction strength. We reiterate here that in the present work we
neglect all such effects. We include the electron-electron interactions
within the self-consistent mean-field approximation, which in the
first order in the electric field gives the density matrix correction
that is proportional to the density response,

\begin{eqnarray*}
 & \left[H_{ee},\hat{g}_{\mathbf{k,k+q}}\right]=V_{q}\rho^{(10)}(q)\left(\hat{g}_{\mathbf{k+q}}^{(0)}-\hat{g}_{\mathbf{k}}^{(0)}\right)=0,
\end{eqnarray*}
which therefore vanishes. As a result, the interactions produce no
change in Eq.\ (\ref{eq:g10}) within the chosen approximation scheme
in our model. The full inclusion of interaction-induced self-energy
and vertex corrections in the graphene dynamical conductivity in the
doped situation in the presence of impurity disorder is a formidable
task, which is well beyond the scope of our work where our interest
lies in treating disorder and interaction on an equal footing, treating
disorder perturbatively (i.e., high-frequency approximation) and interaction
at a mean-field level (i.e., RPA, which correctly incorporates the
plasmon collective mode in the high-frequency conductivity).

The static density fluctuation induced by the disorder potential $\hat{g}_{\mathbf{k,k+q}}^{(01)}$
satisfies Eq.\ (\ref{eq:EOMomegaDiag}) with the first-order terms
in disorder included and the terms proportional to the electric field
omitted in the right-hand side,

\begin{eqnarray}
 & \hat{g}_{\mathbf{k,k+q}}^{(01)}(0)=\sum_{ss'}\frac{\mathcal{P}_{s\mathbf{n_{k}}}\left[\hat{g}_{\mathbf{k+q}}^{(0)}-\hat{g}_{\mathbf{k}}^{(0)}\right]\mathcal{P}_{s'\mathbf{n_{k+q}}}}{s'\epsilon_{\mathbf{k+q}}-s\epsilon_{\mathbf{k}}}\nonumber \\
 & \times\left[V_{\mathbf{q}}^{(i)}+V_{q}\rho^{(01)}(-\mathbf{q},0)\right],\label{eq:g(01)}
\end{eqnarray}
where $s,s'=\pm1$, and we used Eqs.\ (\ref{eq:[g,Himp]})\ and\ (\ref{eq:[g,Hee]}).
Taking the trace over the spinor indices and summing over $\mathbf{k}$
on both sides of Eq.\ (\ref{eq:g(01)}) we solve it for the density
$\rho^{(01)}(q,0)\equiv\sum_{\mathbf{k}}\mathrm{Tr}\hat{g}_{\mathbf{k,k+q}}^{(01)}(0)$,

\begin{eqnarray}
 & \rho^{(01)}(q,0)=\sum_{\mathbf{k}}\mathrm{Tr}\{\hat{g}_{\mathbf{k,k+q}}^{(01)}(0)\}=\frac{V_{-\mathbf{q}}^{(i)}\chi(q,0)}{\varepsilon(q,0)}.\label{eq:n(01)}
\end{eqnarray}
Here we introduced the RPA dielectric function, $1/\varepsilon(q,\omega)\equiv1/(1-V_{-q}\chi(q,\omega))$,
with the free electron polarization operator defined as\ \cite{hwang_dielectric_2007,wunsch_dynamical_2006},

\begin{eqnarray}
 & \chi(q,\omega)\equiv\eta\sum_{ss',\mathbf{k}}\frac{\left(f_{s'\epsilon_{\mathbf{k+q}}}-f_{s\epsilon_{\mathbf{k}}}\right)\left[1+ss'\mathbf{n_{k}}\cdot\mathbf{n_{k+q}}\right]}{-\omega-s\epsilon_{\mathbf{k}}+s'\epsilon_{\mathbf{k+q}}},\label{eq:Polarization}
\end{eqnarray}
where $\eta=2$ is the spin degeneracy. Going back to Eq.\ (\ref{eq:g(01)})
we obtain the first order correction to the density matrix due to
the disorder potential screened by the interacting electron gas, 
\begin{eqnarray}
 & \hat{g}_{\mathbf{k,k+q}}^{(01)}(0)=\frac{V_{-\mathbf{q}}^{(i)}}{\varepsilon(q,0)}\sum\frac{\left(f_{s'\epsilon_{\mathbf{k+q}}}-f_{s\epsilon_{\mathbf{k}}}\right)\mathcal{P}_{s\mathbf{n_{k}}}\mathcal{P}_{s'\mathbf{n_{k+q}}}}{-s\epsilon_{\mathbf{k}}+s'\epsilon_{\mathbf{k+q}}},\label{eq:Resultg(01)}
\end{eqnarray}
where the sum is over $s,s'=\pm1$.

\subsection{Second order terms\label{sub:Second-order-terms}}

The contribution to the static density fluctuations in the second
order in disorder $\hat{g}_{\mathbf{k,k+q}}^{(02)}$ vanishes after
the averaging over the disorder configurations. This is because the
disorder averaged value $\overline{\left(\hat{g}_{\mathbf{k,k+q}}^{(02)}\right)}=\xi\delta_{q,0}$
is homogeneous in space. The charge neutral distribution of disorder
does not give rise to non-zero homogeneous corrections to the density,
therefore $\xi=0$. This conclusion can also be confirmed by showing
that in the second order in the disorder potential Eq.\ (\ref{eq:g(01)})
does not have non-trivial static homogeneous solutions.

The interplay of the disorder potential and the electric field gives
rise to the dynamical fluctuation of the electron density matrix $\hat{g}_{\mathbf{k,k+q}}^{(11)}(\omega)$
that satisfies the following equation obtained from Eq.\ (\ref{eq:EOMomegaDiag})
in the respective expansion,

\begin{eqnarray}
 & \hat{g}_{\mathbf{k,k+q}}^{(11)}(\omega)=\sum_{ss'}\frac{ie\mathbf{E}\cdot\mathcal{P}_{s\mathbf{n_{k}}}\nabla_{\mathbf{k}}\hat{g}_{\mathbf{k,k+q}}^{(01)}(0)\mathcal{P}_{s'\mathbf{n_{k+q}}}}{s'\epsilon_{\mathbf{k+q}}-s\epsilon_{\mathbf{k}}-\omega}\nonumber \\
 & +\frac{V_{-\mathbf{q}}^{(i)}}{\varepsilon(q,0)}\sum_{ss'}\frac{\mathcal{P}_{s\mathbf{n_{k}}}\left(\hat{g}_{\mathbf{k+q}}^{(10)}(\omega)-\hat{g}_{\mathbf{k}}^{(10)}(\omega)\right)\mathcal{P}_{s'\mathbf{n_{k+q}}}}{s'\epsilon_{\mathbf{k+q}}-s\epsilon_{\mathbf{k}}-\omega}\nonumber \\
 & +V_{q}\rho^{(11)}(\mathbf{q},\omega)\sum_{ss'}\frac{\mathcal{P}_{s\mathbf{n_{k}}}\left(\hat{g}_{\mathbf{k+q}}^{(0)}-\hat{g}_{\mathbf{k}}^{(0)}\right)\mathcal{P}_{s'\mathbf{n_{k+q}}}}{s'\epsilon_{\mathbf{k+q}}-s\epsilon_{\mathbf{k}}-\omega}.\label{eq:g(11)}
\end{eqnarray}
We take the trace over the spinor indices and sum over $k$ on both
sides of Eq.\ (\ref{eq:g(11)}) as before, and solve this equation
for the density fluctuation $\rho^{(11)}(\mathbf{q},\omega)\equiv\sum_{\mathbf{k}}\mathrm{Tr}\hat{g}_{\mathbf{k,k+q}}^{(11)}(\omega)$,

\begin{eqnarray}
 & \rho^{(11)}(\mathbf{q},\omega)=\frac{1}{\varepsilon(q,\omega)}\sum\mathrm{Tr}\left[\frac{ie\mathbf{E}\cdot\mathcal{P}_{s\mathbf{n_{k}}}\nabla_{\mathbf{k}}\hat{g}_{\mathbf{k,k+q}}^{(01)}(0)\mathcal{P}_{s'\mathbf{n_{k+q}}}}{s'\epsilon_{\mathbf{k+q}}-s\epsilon_{\mathbf{k}}-\omega}\right.\nonumber \\
 & \left.+\frac{V_{-\mathbf{q}}^{(i)}}{\varepsilon(q,0)}\frac{\mathcal{P}_{s\mathbf{n_{k}}}\left(\hat{g}_{\mathbf{k+q}}^{(10)}(\omega)-\hat{g}_{\mathbf{k}}^{(10)}(\omega)\right)\mathcal{P}_{s'\mathbf{n_{k+q}}}}{s'\epsilon_{\mathbf{k+q}}-s\epsilon_{\mathbf{k}}-\omega}\right],\label{eq:Rho(11)}
\end{eqnarray}
where the sum is taken over $s,s'=\pm1$ and the momentum $\mathbf{k}$.
Note that the combination of the left-hand side and the last term
on the right-hand side of Eq.\ (\ref{eq:g(11)}) ensure that the
density fluctuation $\rho^{(11)}(\mathbf{q},\omega)$ is proportional
to the inverse of the dynamical dielectric function $1/\varepsilon(q,\omega)$.
Using the explicit result for $\rho^{(11)}(\mathbf{q},\omega)$ we
can find the density matrix correction $\hat{g}_{\mathbf{k,k+q}}^{(11)}(\omega)$.

The full solution $\hat{g}_{\mathbf{k,k+q}}^{(11)}(\omega)$ is straightforward
to obtain. The result however is rather cumbersome (and not particularly
illuminating) and therefore below we will present only the terms that
play a key role in the plasmon enhanced scattering rate which dominates
the frequency-dependent optical response at low frequencies $1/\tau\ll\omega\ll E_{F}$.

\subsection{Electric current }

Using Eq.\ (\ref{eq:EOMomegaDiag}) we can express the linear response
current in terms of $\hat{g}_{\mathbf{k,k+q}}^{(11)}(\omega),$ which
spares us the necessity of calculating $\hat{g}_{\mathbf{k,k+q}}^{(12)}(\omega)$.
We multiply Eq.\ (\ref{eq:EOMomegaDiag}) by $\Sigma^{j}$ (where
$j=x,y$) take the trace in the spinor space and sum over $k$. Comparing
the result to the definition in Eq.\ (\ref{eq:CurrentGeneral}) we
arrive at,

\begin{eqnarray}
 & J_{total}^{j}=ev\sum_{ss',\mathbf{k}}\frac{\mathrm{Tr}\left\{ \Sigma^{j}\mathcal{P}_{s\mathbf{n_{k}}}\left[H_{ee}+H_{imp},\hat{g}_{\mathbf{k,k}}(\omega)\right]\mathcal{P}_{s'\mathbf{n_{k}}}\right\} }{(s'-s)\epsilon_{\mathbf{k}}-\omega}\nonumber \\
 & +ev\sum_{ss',\mathbf{k}}\frac{\mathrm{Tr}\left\{ \Sigma^{j}\mathcal{P}_{s\mathbf{n_{k}}}\left(ie\mathbf{E}\cdot\nabla_{\mathbf{k}}\hat{g}_{\mathbf{k,k}}(\omega)\right)\mathcal{P}_{s'\mathbf{n_{k}}}\right\} }{(s'-s)\epsilon_{\mathbf{k}}-\omega}.\label{eq:Current}
\end{eqnarray}
The term in the second line in Eq.\ (\ref{eq:Current}) is proportional
to the homogeneous component of the correlator $\hat{g}_{\mathbf{k,k}}$
and therefore does not give any contribution to the disorder induced
correction, see Sec.\ \ref{sub:Second-order-terms} for details.
The presence of this free electron contribution is an artifact of
the perturbation theory, which does not correctly describe the broadening
of the singular response at $\omega=0$. In the following we focus
on the high-frequency correction to the optical conductivity due to
disorder scattering, which is accurately described by the first line
in Eq.\ (\ref{eq:Current}), and is denoted $J^{j}(\omega)$ in the
following. Using Eqs.\ (\ref{eq:[g,Himp]}), (\ref{eq:[g,Hee]})
and the cyclic property of trace we write the current response due
to the disorder scattering accurate in the second order in disorder
as,

\begin{eqnarray}
 & J^{j}(\omega)=ev\sum\frac{\mathrm{Tr}\left\{ \mathcal{P}_{s'\mathbf{n_{k}}}\Sigma^{j}\mathcal{P}_{s\mathbf{n_{k}}}\left(\hat{g}_{\mathbf{k-q,k}}(\omega)-\hat{g}_{\mathbf{k,k+q}}(\omega)\right)\right\} }{(s'-s)\epsilon_{\mathbf{k}}-\omega}\nonumber \\
 & \times\left[V_{\mathbf{q}}^{(i)}+V_{q}\rho(-\mathbf{q},\omega)\right],\label{eq:Current-Step1}
\end{eqnarray}
where the sum is taken over $s,s'=\pm1$ and the momenta $\mathbf{k}$
and $\mathbf{q}$. The numerator in Eq.\ (\ref{eq:Current-Step1})
contains the bracket $\left(\hat{g}_{\mathbf{k-q,k}}(\omega)-\hat{g}_{\mathbf{k,k+q}}(\omega)\right)$.
We shift the sum over momentum $\mathbf{k\rightarrow k+q}$ in the
first term of this bracket so that the corresponding difference appears
in the factor in front of the bracket denoted $\Xi^{j}$,

\begin{eqnarray}
 & J^{j}(\omega)=ev\sum_{ss',\mathbf{k,q}}\mathrm{Tr}\left\{ \Xi_{ss'}^{j}\hat{g}_{\mathbf{k,k+q}}(\omega)\right\} \nonumber \\
 & \times\left[V_{\mathbf{q}}^{(i)}+V_{q}\rho(-\mathbf{q},\omega)\right],\label{eq:Current-Step2}
\end{eqnarray}
where we introduced,

\begin{eqnarray*}
 & \Xi_{ss'}^{j}=\frac{\mathcal{P}_{s'\mathbf{n_{k+q}}}\Sigma^{j}\mathcal{P}_{s\mathbf{n_{k+q}}}}{-\omega+(s'-s)\epsilon_{\mathbf{k+q}}}-\frac{\mathcal{P}_{s'\mathbf{n_{k}}}\Sigma^{j}\mathcal{P}_{s\mathbf{n_{k}}}}{-\omega+(s'-s)\epsilon_{\mathbf{k}}}.
\end{eqnarray*}
This procedure has to be formally justified by an introduction of
a band cut off. The details of this formulation are described in Appendix\ \ref{sec:Appendix-I}.
Finally, using the expansion\ (\ref{eq:g-expansion}) and the result\ (\ref{eq:n(01)})
the current response Eq.\ (\ref{eq:Current-Step2}) to the external
electric field in the second order in the screened disorder potential
is written as,

\begin{eqnarray}
 & J^{j}(\omega)=ev\sum_{ss',k}\textrm{Tr}\left\{ \Xi_{ss'}^{j}\left(\frac{V_{\mathbf{q}}^{(i)}}{\varepsilon(q,0)}\hat{g}_{\mathbf{k,k+q}}^{(11)}(\omega)\right.\right.\nonumber \\
 & \left.\left.+V_{q}\rho^{(11)}(-\mathbf{q},\omega)\hat{g}_{\mathbf{k,k+q}}^{(01)}(0)\right)\right\} .\label{eq:CurrentPert}
\end{eqnarray}

\section{Plasmon propagator in graphene\label{sec:Plasmon-propagator}}

The excitation spectrum of the interacting Dirac electron gas consists
of the electron-hole continuum and the collective plasmon mode. The
latter is described by the pole of the inverse dielectric function
given within RPA by\ \cite{hwang_dielectric_2007,wunsch_dynamical_2006},

\begin{equation}
1-V_{q}\chi(q,\omega)=0.\label{eq:RPApole}
\end{equation}
Here $\chi(q,\omega)$ is the free electron polarization function
Eq.\ (\ref{eq:Polarization}). Equation\ (\ref{eq:RPApole}) has
solutions only when $\omega>vq$, which correspond to the plasmon
band with the dispersion\ \cite{hwang_dielectric_2007,wunsch_dynamical_2006},
\[
\omega_{pl}=v\sqrt{\eta\alpha k_{F}}\sqrt{q}.
\]
In this frequency range $\omega>vq$, the imaginary part of the dynamical
dielectric function is proportional to the plasmon propagator, 
\begin{eqnarray}
 & \Imm \frac{\chi(q,\omega)}{\varepsilon(q,\omega)}=\Imm D_{pl}(q,\omega),\ vq<\omega.\label{eq:DielSplit}
\end{eqnarray}
In the relativistic electron gas the Landau damping is absent, i.e.
$\Imm\chi(q,\omega)=0$, in the frequency range $vq<\omega<2E_{F}-vq$\ \cite{wunsch_dynamical_2006,hwang_dielectric_2007}.
Therefore within this range we can approximate the imaginary part
of the plasmon propagator by a $\delta$-function neglecting the relatively
weak damping by disorder, 
\begin{eqnarray}
 & \Imm\frac{1}{\varepsilon(q,\omega)}\approx\frac{1}{V_{q}\frac{\partial\chi(q,\omega_{pl}(q))}{\partial\omega}}\pi\delta(\omega-\omega_{pl}(q)).\label{eq:PlasmonPole}
\end{eqnarray}
The plasmon momentum is restricted to $vq<\omega$ which therefore
allows an expansion in both the frequency $\omega\ll E_{F}$ and momentum
$q\ll k_{F}$ in the prefactor in front of the plasmon propagator
in Eq.\ (\ref{eq:CurrentPert}). At higher frequencies, when the
condition $\omega+vq>2E_{F}$ is satisfied, the plasmon is damped
by electron-hole excitations, and the frequency-dependent plasmon
width (i.e. Landau damping) has to be taken into account. The plasmon
damping does not give rise to any qualitatively different behavior.
Therefore we consider only the $\delta$-function form of the plasmon
propagator which gives quantitatively accurate results at frequencies
$\omega\ll E_{F}$. Electron-hole excitations determine the behavior
of the dielectric function for $\omega<vq$,

\begin{eqnarray}
 & \Imm\frac{1}{\varepsilon(q,\omega)}=\frac{\Imm\chi(q,\omega)}{|\varepsilon(q,\omega)|^{2}}.\label{eq:Dielelectron-hole}
\end{eqnarray}
At low frequencies $\omega\ll E_{F}$ the frequency dependence of
the dynamical screening due to electron-hole excitations given by
Eq.\ (\ref{eq:Dielelectron-hole}) is weak. Therefore the frequency
dependence of the current response is dominated by the terms proportional
to the plasmon propagator in Eq.\ (\ref{eq:PlasmonPole}). This allows
us to drop the frequency dependence in all other terms in the current
response,

\[
J^{j}(\omega)\approx J^{j}(\omega=0)+J_{pl}^{j}(\omega),
\]
where $J_{pl}^{j}(\omega)$ is the frequency-dependent part of Eq.\ (\ref{eq:CurrentPert})
that contains the plasmon propagator,

\begin{eqnarray}
 & J_{pl}^{j}(\omega)=ev\sum\textrm{Tr}\left\{ \Xi_{ss'}^{j}\left(V_{q}\rho^{(11)}(-\mathbf{q},\omega)\hat{g}_{\mathbf{k,k+q}}^{(01)}(0)\right.\right.\nonumber \\
 & \left.\left.+\frac{V_{\mathbf{q}}^{(i)}V_{q}\rho^{(11)}(\mathbf{q},\omega)}{\varepsilon(q,0)}\sum\frac{\left(f_{t'\epsilon_{\mathbf{k+q}}}-f_{t\epsilon_{\mathbf{k}}}\right)\mathcal{P}_{t\mathbf{n_{k}}}\mathcal{P}_{t'\mathbf{n_{k+q}}}}{t'\epsilon_{\mathbf{k+q}}-t\epsilon_{\mathbf{k}}-\omega}\right)\right\} ,\label{eq:Jpl}
\end{eqnarray}
where we kept only the last term in Eq.\ (\ref{eq:g(11)}) for $\hat{g}_{\mathbf{k,k+q}}^{(11)}(\omega),$
which is the only one containing the plasmon propagator. The sum in
the first line of Eq.\ (\ref{eq:Jpl}) is over $s,s'=\pm1$ and the
momenta $\mathbf{k}$ and $\mathbf{q}$, in the second line the sum
is over $t,t'=\pm1$.

The final analytical expression for the plasmon enhanced optical conductivity
is rather cumbersome and therefore is left for the Appendix\ \ref{sub:Full-solution}.
In the next section we find a simplified form of this result introducing
an additional expansion parameter, $\omega/E_{F}\ll1$.

\section{Asymptotic solutions\label{sec:Asymptotic-solutions}}

\subsection{Intraband solution of the equation of motion }

At frequencies much lower than the Fermi level $\omega/E_{F}\ll1$
both the static scattering rate and the frequency-dependent plasmon-induced
correction to it are dominated by the intraband processes. Therefore
in the right-hand side of Eq.\ (\ref{eq:EOMomegaDiag}) we can neglect
all interband terms that contain large denominators $\approx2E_{F}$
in the low-frequency limit,

\begin{eqnarray}
 & \hat{g}_{\mathbf{k,k+q}}(\omega)\approx\frac{\mathcal{P}_{\mathbf{n_{k}}}\left[H_{ee}+H_{imp},\hat{g}_{\mathbf{k,k+q}}(\omega)\right]\mathcal{P}_{\mathbf{n_{k+q}}}}{\epsilon_{\mathbf{k+q}}-\epsilon_{\mathbf{k}}-\omega}\nonumber \\
 & +\frac{\mathcal{P}_{\mathbf{n_{k}}}\left(ie\mathbf{E}\cdot\nabla_{\mathbf{k}}\hat{g}_{\mathbf{k,k+q}}(\omega)\right)\mathcal{P}_{\mathbf{n_{k+q}}}}{\epsilon_{\mathbf{k+q}}-\epsilon_{\mathbf{k}}-\omega}.\label{eq:ProjectedEOM}
\end{eqnarray}
Using Eq.\ (\ref{eq:ProjectedEOM}) instead of Eq.\ (\ref{eq:EOMomegaDiag})
greatly simplifies the algebra needed to implement the program outlined
in Sec.\ \ref{sec:EOMSolution}. Thus solving Eq.\ (\ref{eq:ProjectedEOM})
for $\hat{g}_{\mathbf{k,k+q}}^{(01)}(0)$ and $\rho^{(11)}(\mathbf{q},\omega)$
and using Eq.\ (\ref{eq:CurrentPert}) and performing some tedious
but straightforward algebraic calculations we arrive at,

\begin{equation}
\Ree\sigma=\Ree\sigma_{D}+\Ree\sigma_{pl},\label{eq:SigmaSum}
\end{equation}
where,

\begin{eqnarray}
 & \Ree\sigma_{D}=-\frac{e^{2}v^{2}\rho_{imp}}{2\omega^{2}}\sum_{q}\frac{V_{q}^{2}e^{-2qd}}{\varepsilon^{2}(q,0)}\Imm\Phi_{\sigma}(q,\omega),\label{eq:SigmaDrude}\\
 & \Ree\sigma_{pl}=-\frac{e^{2}v^{2}\rho_{imp}}{2\omega}\sum_{q}\frac{V_{q}^{3}e^{-2qd}}{\varepsilon^{2}(q,0)}\Imm\frac{\left[X_{j}(q,\omega)\right]^{2}}{\varepsilon(q,\omega)},\label{eq:SigmaPlasmon}
\end{eqnarray}
and $j=x,y$, and we used Eq.~(\ref{VCorr}) for the averaged correlator
of the disorder potential. We introduced above the correlation functions,

\begin{eqnarray}
 & \Phi_{\sigma}(q,\omega)\equiv\eta\sum_{k}\frac{\left(f_{\epsilon_{\mathbf{k+q}}}-f_{\epsilon_{\mathbf{k}}}\right)\left[\mathbf{n_{k+q}}-\mathbf{n_{k}}\right]^{2}\mathcal{F}_{\mathbf{k,k+q}}}{\left(-\omega-i\delta-\epsilon_{\mathbf{k}}+\epsilon_{\mathbf{k+q}}\right)\left(-\epsilon_{\mathbf{k}}+\epsilon_{\mathbf{k+q}}\right)},\label{eq:chiSigma}
\end{eqnarray}
and

\begin{eqnarray}
 & X_{j}(q,\omega)\equiv\eta\sum_{k}\frac{\left(f_{\epsilon_{\mathbf{k+q}}}-f_{\epsilon_{\mathbf{k}}}\right)\left[n_{\mathbf{k+q}}^{j}-n_{\mathbf{k}}^{j}\right]\mathcal{F}_{\mathbf{k,k+q}}}{\left(-\omega-i\delta-\epsilon_{\mathbf{k}}+\epsilon_{\mathbf{k+q}}\right)\left(-\epsilon_{\mathbf{k}}+\epsilon_{\mathbf{k+q}}\right)}.\label{eq:Xj}
\end{eqnarray}
Here we introduced a notation,

\begin{equation}
\mathcal{F}_{\mathbf{k,k+q}}\equiv\left(1+\mathbf{n_{k}\cdot\mathbf{n_{k+q}}}\right).\label{eq:PhaseFactor}
\end{equation}
The first term in Eq.\ (\ref{eq:SigmaSum}) reproduces the high-frequency
expansion of the standard Drude conductivity,

\begin{equation}
\Ree\sigma{}_{Drude}=\frac{\eta e^{2}\nu v^{2}\tau}{1+(\omega\tau)^{2}}\approx\frac{\eta e^{2}\nu v^{2}}{\omega^{2}\tau},\label{eq:DrudeForm}
\end{equation}
with the frequency-independent transport relaxation rate due to impurity
scattering. Here $\nu$ is the density of states of the Dirac electrons
per spin and valley. To demonstrate the relation between Eqs.\ (\ref{eq:SigmaDrude})
and\ (\ref{eq:DrudeForm}) explicitly we keep only the main term
in the low-frequency expansion of Eq.\ (\ref{eq:chiSigma}),

\begin{eqnarray*}
 & \Imm\Phi_{\sigma}(q,\omega)\approx q^{2}\Imm\frac{\chi(q,\omega)-\chi(q,0)}{\omega}.
\end{eqnarray*}
Substituting the above result into Eq.\ (\ref{eq:SigmaDrude}) and
comparing it to Eq.\ (\ref{eq:DrudeForm}) we reproduce the standard
definition\ \cite{das_sarma_electronic_2011} of the static elastic
scattering rate due to screened Coulomb disorder in graphene,

\begin{eqnarray}
 & \Ree\sigma{}_{D}\approx\frac{2e^{2}\nu v^{2}}{\omega^{2}\tau(0)},\nonumber \\
 & \frac{1}{\tau(0)}\nonumber \\
 & =\frac{2\pi\rho_{imp}}{\hbar}\sum_{\mathbf{p}}\frac{e^{-2|\mathbf{p-\tilde{p}}|d}V_{|\mathbf{p-\tilde{p}}|}^{2}\mathcal{F}_{\mathbf{p},\mathbf{\tilde{p}}}(1-\cos\theta_{\mathbf{p\tilde{p}}})\delta(\epsilon_{\mathbf{p}}-\epsilon_{\mathbf{\tilde{p}}})}{\varepsilon^{2}(|\mathbf{p-\tilde{p}}|,0)},\label{eq:StandartRate}
\end{eqnarray}
where we restore the Planck's constant.

The second term in Eq.\ (\ref{eq:SigmaSum}) describes the frequency-dependent
contribution to the optical conductivity due to the plasmon anomaly.
This part can be simplified by keeping only the main order in in the
parameter $\omega/E_{F}\ll1$ in Eq.\ (\ref{eq:Xj}),

\begin{eqnarray*}
 & X_{j}(q,\omega)\approx2\eta\frac{1}{v}\int\frac{d\varphi}{(2\pi)^{2}}\frac{q^{j}+n_{k}^{j}\mathbf{n_{k}}\cdot\mathbf{q}}{\omega-v\mathbf{n_{k}\cdot q}}.
\end{eqnarray*}
Taking the integral over the angle $\varphi$ we arrive at,

\begin{eqnarray}
 & X_{j}(q,\omega)=\eta\frac{1}{\pi}\frac{n_{q}^{j}}{v^{2}}\frac{\Omega}{\sqrt{1-\Omega^{2}}}\left(1+\frac{1}{\sqrt{1-\Omega^{2}}+1}\right)\nonumber \\
 & \approx\frac{\zeta\eta}{\pi}\frac{n_{q}^{j}}{v^{2}}\Omega,\label{eq:XjqExp}
\end{eqnarray}
where $\Omega\equiv\frac{vq}{\omega}$ and $\zeta=3/2.$ Therefore
the expansion of Eq.\ (\ref{eq:SigmaPlasmon}) in the parameter,
$\omega/E_{F}$ with the use of Eq.\ (\ref{eq:XjqExp}) gives, 
\begin{eqnarray}
 & \Ree\sigma_{pl}=-\frac{\zeta^{2}\eta^{2}e^{2}\rho_{imp}}{2\pi^{2}\omega^{3}}\sum_{q}\frac{q^{2}V_{q}^{3}e^{-2qd}}{\varepsilon^{2}(q,0)}\Imm\frac{1}{\varepsilon(q,\omega)}.\label{Sigma}
\end{eqnarray}

\subsection{Expansion of the full solution in the parameter, $\omega/E_{F}\ll1$}

We can verify the intraband result for $\Ree\sigma_{pl}$ obtained above by extracting
the main order contribution in the parameter $\omega/E_{F}\ll1$ from
the plasmon anomaly induced correction to the optical conductivity Eq. (\ref{eq:RPAFull})
obtained by solving the full equation of motion valid at frequencies
$\omega<2E_{F}$. We find that the result of this procedure coincides
with Eq.\ (\ref{Sigma}).

The result\ (\ref{Sigma}) is somewhat analogous to the one derived
for the case of the 2D electron gas with the parabolic dispersion\ \cite{GotzeWolfle72},
albeit with a different numerical coefficient. However, in the case
of graphene Eq.\ (\ref{Sigma}) corresponds to the main order in
$\omega/E_{F}\ll1$ and describes only the frequency-dependent scattering
rate due to the plasmon anomaly, the frequency dependence of screening
due to the electron-hole excitations is neglected. At higher frequencies
$\omega\sim E_{F}$ Eq.\ (\ref{eq:RPAFull}) provides a more quantitatively
accurate description. At yet higher frequencies $\omega\gtrsim2E_{F}$
the solution of Eq.\ (\ref{eq:g(11)}) with\ (\ref{eq:CurrentPert})
has to be used including all interband effects.

\section{Results\label{sec:Results}}

\subsection{High frequency regime: $\omega\tau\gg1$}

\begin{figure}
\includegraphics[width=0.75\columnwidth]{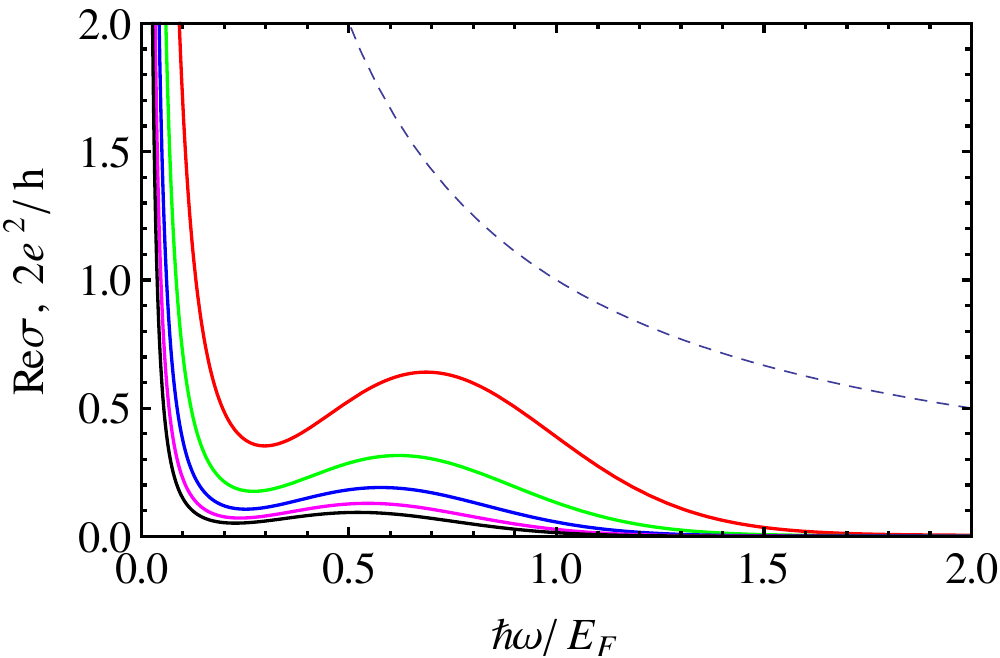}

\caption{\label{fig:conductance}The real part of the conductivity (in units
of conductance quantum) approximated by the sum of the Drude peak
with the frequency-independent scattering rate and the plasmon peak
Eq.\ (\ref{eq:PlasmonPeak}). The Drude peak corresponds to the very
narrow feature at $\omega=0$. The dashed line shows the limit of
applicability of the perturbative expansion $\omega\tau\sim1$. The
impurity density is $\rho_{imp}=6\times10^{12}cm^{-2}$ located at
$d=10\times10^{-9}m$, the electron density varies from $10^{12}cm^{-2}$
to $3\times10^{12}cm^{-2}$ top to bottom on the right, the strength
of interactions is $\alpha=0.58$ corresponding to graphene on hBN.}
\end{figure}

Substituting the plasmon propagator Eq.\ (\ref{eq:PlasmonPole})
into the expression for the optical conductivity Eq.\ (\ref{Sigma})
and taking the trivial integral over the momentum $q,$ which is equivalent
to the substitution $q\rightarrow q_{\omega}\equiv\omega^{2}/(2\alpha vE_{F})$,
we calculate the frequency-dependent part of the optical conductivity
(in units of conductance quantum $\sigma_{0}\equiv2e^{2}/h$),

\begin{eqnarray}
 & \Ree\sigma_{pl}=\frac{9\pi\rho_{imp}\sigma_{0}}{64\alpha^{\frac{1}{2}}\left(k_{F}d\right)^{\frac{3}{2}}\rho}\left(\frac{\hbar\omega}{cE_{F}}\right)^{3}e^{-\left(\frac{\hbar\omega}{cE_{F}}\right)^{2}}.\label{eq:PlasmonPeak}
\end{eqnarray}
with $c\equiv\sqrt{\alpha/(k_{F}d)}$. Here and throughout this section
we restore the Planck's constant. Figure\ \ref{fig:conductance}
shows Eq.\ (\ref{eq:SigmaSum}) using Eq.\ (\ref{eq:PlasmonPeak})
and the Lorentzian peak $\Ree\sigma_{D}(\omega)\approx\Ree\sigma_{Drude}(\omega)$
with the static elastic relaxation rate given by Eq.\ (\ref{eq:StandartRate}).
The effect of the plasmon pole on the optical conductivity is weak
at lower frequencies $\hbar\omega\ll cE_{F}$, however, at higher
frequencies $\hbar\omega\gtrsim cE_{F}$ there is a sharp increase
in the optical conductivity with increasing frequency that is limited
by the flattening of the potential of charged impurities as determined
by their position $d$. A consequence of this behavior is a broad
peak in the frequency-dependent optical conductivity with the maximum
located at, 
\begin{equation}
\hbar\omega^{\ast}=\sqrt{\frac{3}{2}}cE_{F},
\end{equation}
and the width defined by the solution of the equation $\partial_{\omega}^{2}\sigma_{pl}(\omega)=0$,

\begin{equation}
\hbar\delta\omega=\frac{5}{2}cE_{F},
\end{equation}
as shown in Fig.\ \ref{fig:conductance}.

\begin{figure}
\includegraphics[width=0.45\columnwidth]{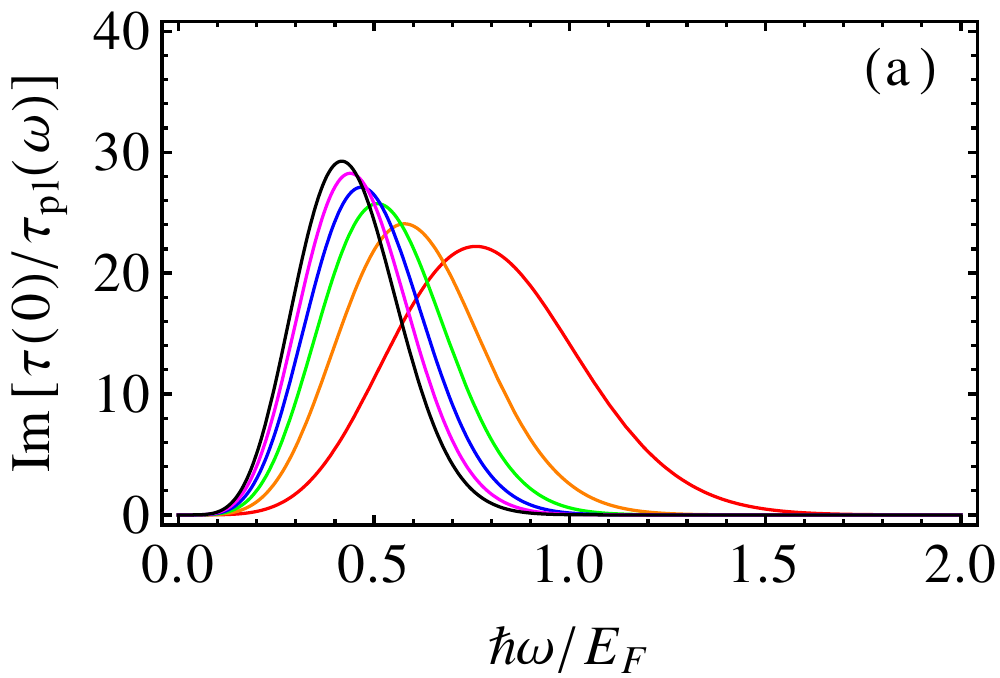}\includegraphics[width=0.465\columnwidth]{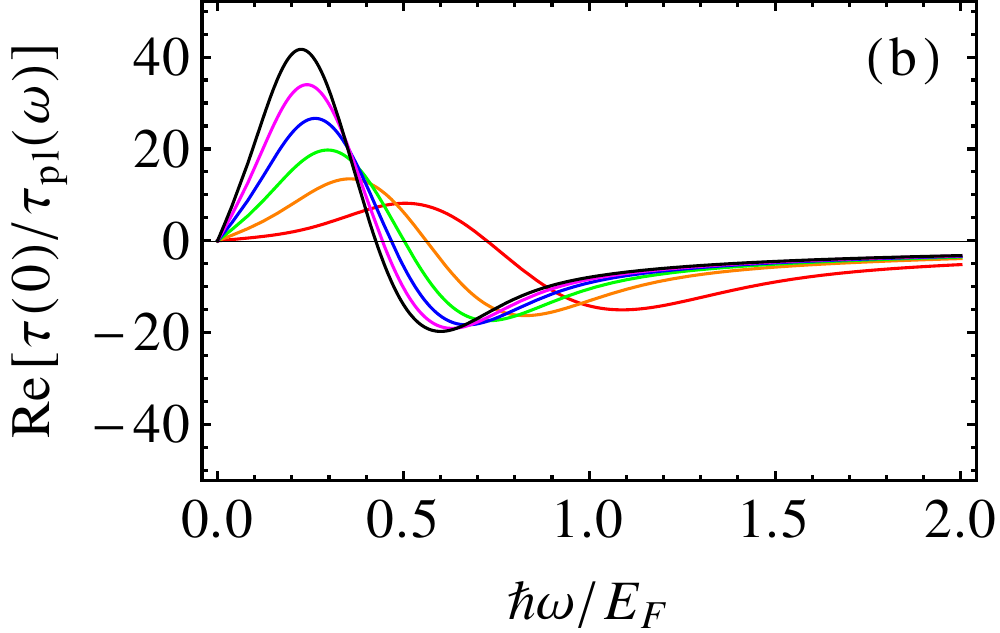}\caption{\label{fig:Rate}The ratio of the static transport scattering time
$\tau(0)$ to the frequency-dependent scattering rate due to plasmon
anomaly $\tau_{pl}(\omega)$ for graphene. The parameters are, $d=20nm$,
the dimensionless interaction strength is $\alpha=0.58$, the impurity
density $\rho_{imp}=3\times10^{12}cm^{-2}$. Different lines correspond
to the different electron densities from $\rho=0.5\times10^{12}cm^{-2}$
to $\rho=5.5\times10^{12}cm^{-2}$ top to bottom on the right. Maximum
of the scattering rate $\hbar\omega_{max}=E_{F}\sqrt{5\alpha/(2k_{F}d)}$
is dependent on the electron density in the system. (a) and (b) correspond
to the imaginary and real part of the complex frequency dependent
scattering rate $1/\tau_{pl}(\omega)$, respectively.}
\end{figure}

We define a frequency-dependent relaxation rate associated with the
plasmon enhanced dissipation using the Drude form Eq.\ (\ref{eq:DrudeForm})
and comparing it to Eq.\ (\ref{eq:PlasmonPeak}) we get, 
\begin{eqnarray}
 & \Imm\frac{\hbar}{\tau_{pl}(\omega)}=\frac{9\pi\rho_{imp}}{64\alpha^{2}\rho}\frac{(\hbar\omega)^{5}}{E_{F}^{4}}e^{-\left(\frac{\hbar\omega}{cE_{F}}\right)^{2}},\label{eq:PlasmonRate}
\end{eqnarray}
In a wide range of parameters the plasmon enhanced scattering rate
varies greatly with frequency and can be well over an order of magnitude
larger than the static transport scattering rate, see Fig.\ \ref{fig:Rate}(a).
This strong frequency dependence of the imaginary part of the relaxation
rate gives rise to a non-zero real part shown in Fig.\ \ref{fig:Rate}(b).
The strong dynamical variation of the scattering rate leads to the
spectral weight redistribution in the optical conductivity which will
be particularly important in the strongly disordered samples at low
densities for which the above perturbation theory is expected to fail.
This regime will be discussed in the following section.

In the following we take into account the finite spread of the spatial
distribution of impurities along the $z$ axis perpendicular to the
graphene layer. Including the distribution spread requires the additional
averaging of Eq.\ (\ref{eq:PlasmonPeak}) over positions of impurities
along the $z$ axis, which results in an overall suppression of the
magnitude of the effect of the plasmon-enhanced scattering. For comparison,
we calculate the plasmon-enhanced frequency-dependent optical conductivity
for the case of a uniform density of impurities along $Oz$ in a layer
$d_{1}\leq z\leq d_{2}$, so that the 3D impurity density equals $\rho_{imp}^{3D}=\frac{\rho_{imp}}{|d_{2}-d_{1}|}$.
The result for $d_{2}\gg d_{1}$ reads,

\begin{eqnarray}
 & \Ree\sigma_{pl}^{3D}\approx\frac{1}{k_{F}|d_{2}-d_{1}|}\frac{9\pi\rho_{imp}\sigma_{0}}{64\alpha^{\frac{1}{2}}(k_{F}d_{1})^{\frac{1}{2}}\rho}\frac{\hbar\omega}{c_{1}E_{F}}e^{-\left(\frac{\hbar\omega}{c_{1}E_{F}}\right)^{2}}.\label{eq:3DPeak}
\end{eqnarray}
where $c_{1}=\sqrt{\alpha/(k_{F}d_{1})}$. Note that the 3D impurity
distribution is reflected in a different power in the frequency-dependent
optical conductivity and therefore in a different shape of the plasmon-induced
peak. The maximum of the peak given by Eq.\ (\ref{eq:3DPeak}) is
suppressed with respect to the peak maximum of Eq.\ (\ref{eq:PlasmonPeak})
with $d=d_{1}$ by 
\[
\frac{\max\left[\Ree\sigma_{pl}^{3D}\right]}{\max\left[\Ree\sigma_{pl}\right]}\approx4.7\times\frac{\alpha d_{1}}{|d_{2}-d_{1}|},
\]
which is not necessarily a very small number for typical parameters.
Therefore we expect the effect of plasmon-enhanced scattering to be
observable even in the case of the impurities distributed three dimensionally
throughout the substrate layer in the typical field effect geometry
of the graphene based devices.

\subsection{Low frequency regime: $\omega\tau\lesssim1$\label{sub:Low-frequency-regime:}}

In the following we extend the perturbative results (relying on $\omega\tau\gg1$)
obtained above into the low-frequency regime, $\omega\tau\lesssim1$.
This procedure allows us to make a qualitative prediction about the
low-density behavior of strongly disordered samples. In this low-frequency
regime the spectral weight redistribution in the optical conductivity
is expected to be substantial and the shape of the Drude peak is expected
to be non-Lorentzian.

The frequency-dependent scattering is described by a generalized ``self-energy''
or memory function $M(\omega)$ that reflects the divergent nature
of the response of the electron gas at $\omega=0$\ ,

\begin{eqnarray*}
 & \sigma_{M}(\omega)=\frac{i\sigma_{0}E_{F}}{\omega+M(\omega)}.
\end{eqnarray*}
The real part of the optical conductivity then reads,

\begin{eqnarray*}
 & \Ree\sigma_{M}=\frac{1}{\omega^{2}}\frac{\sigma_{0}E_{F}\Imm M}{\left(1+\frac{\Ree M}{\omega}\right)^{2}+\left(\frac{\Imm M}{\omega}\right)^{2}}.
\end{eqnarray*}
In the limit $\frac{\Ree M}{\omega}\ll1$ and $\frac{\Imm M}{\omega}\ll1$
we can match the expression for the memory function with the perturbative
solution found above,

\begin{eqnarray}
 & \Imm M\approx\frac{\omega^{2}}{\sigma_{0}E_{F}}\Ree\sigma(\omega),\label{eq:MemoryMatching}
\end{eqnarray}
where the right-hand side is given by Eq.\ (\ref{eq:SigmaSum}).
On the other hand at $\omega=0$ the scattering rate in the perturbative
expression for the optical conductivity in Eq.\ (\ref{eq:SigmaSum})
coincides with the static elastic transport scattering rate Eq.\ (\ref{eq:StandartRate}).
Note that the contribution due to the plasmon anomaly Eq.\ (\ref{eq:PlasmonPeak})
vanishes with $\omega\rightarrow0$. The memory function is therefore
a function of frequency that equals $\Imm M(0)=\frac{1}{\tau(0)}$
at $\omega=0$ and is given by Eq.\ (\ref{eq:MemoryMatching}) at
$\omega\tau\gg1$. Therefore using the expansion\ (\ref{eq:MemoryMatching})
with the right hand side given by Eq.\ (\ref{eq:SigmaSum}) in the
whole range of frequencies $0\leq\hbar\omega\ll E_{F}$ provides a
reasonable interpolation for the behavior of the low-frequency optical
conductivity. This interpolation would fail in the presence of any
divergence in the memory function at low frequencies $\omega\tau\lesssim1$.
We do not anticipate such a divergence at frequencies of interest
here $\hbar\omega\ll2E_{F}$ (where the interband processes are not
expected to play a role).

\begin{figure}
\includegraphics[width=0.75\columnwidth]{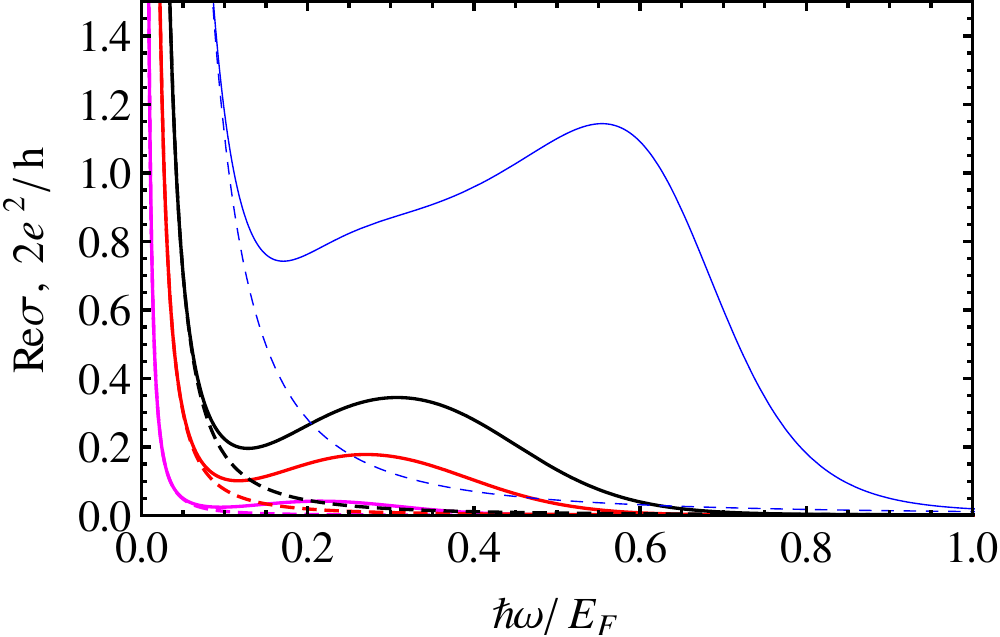}

\caption{\label{fig:Memory}The real part of the conductivity extrapolated
to low frequencies, see Sec \ref{sub:Low-frequency-regime:}. The
parameters are $\alpha=0.15,\rho_{imp}=6\times10^{12}\textrm{c\ensuremath{m^{-2}}},d=10\times10^{-9}\textrm{m}$
with the electron density changing from $\rho=10^{12}\textrm{c\ensuremath{m^{-2}}}$
to $\rho=4\times10^{12}\textrm{c\ensuremath{m^{-2}}}$ for different
lines, top to bottom. Dashed lines correspond to the real part of
Drude conductivity with the frequency-independent scattering rate. }
\end{figure}

The result of the low-frequency interpolation procedure is presented
in Fig.\ \ref{fig:Memory}, which shows the strong evolution of the
Drude peak shape with the changing density (which corresponds to the
different lines on the plot). In particular, the top (blue) line in
Fig.\ \ref{fig:Memory} demonstrates the importance of the non-zero
real part of the memory function, which gives rise to the strong redistribution
of the spectral weight in the optical conductivity and the resulting
non-trivial shape of the tail of the Drude peak.

\section{Discussion\label{sec:Discussion}}

In summary, we have developed an equation of motion approach to describe
the linear response of Dirac electrons in the presence of electron-electron
interaction and disorder scattering due to charged impurities. We
obtain quantitative predictions for the optical conductivity in the
high-frequency regime $\omega\tau\gg1$. We show that the presence
of a plasmon pole in the dynamical dielectric function leads to a
strong enhancement of the dissipation at finite frequency reflected
in a wide peak in the real part of the optical conductivity. Characteristics
of the predicted peak feature are strongly dependent on the strength
of the electron-electron interaction, the location of the impurities
with respect to the 2D Dirac electron gas, and can be tuned by tuning
the Fermi level by changing the carrier density. Extrapolating to
the low-frequency regime $\omega\tau\lesssim1$ we find that the plasmon
enhanced scattering may determine the non-Lorentzian shape of the
Drude response.

The theory developed here is quantitatively accurate only in the high-frequency
regime $\omega\tau\gg1$. An accurate description at low frequencies
must be based on the solution of the Bethe-Salpeter equation for the
vertex function or some analog of it, which is beyond the scope of
the current work where we focus on the interplay between disorder
and interaction perturbatively (thus necessitating the high-frequency
perturbative expansion). We have, however, provided a low-frequency
extrapolation of our theory, which should have qualitative, if not
quantitative, validity, as long as the relevant memory function does
not have a singularity at low frequencies. This extrapolation is really
an interpolation based on an extension of our high-frequency theory
to lower frequencies by using the known zero-frequency Drude result.

We have neglected effects of the many-body renormalization of the
electronic spectrum and exciton energies by electron-electron interactions.
Within the parameter range considered in this paper, high Fermi level
and relatively low frequencies $\hbar/\tau\ll\hbar\omega\ll E_{F}$
we expect the renormalization effects to give only a small quantitative
rather than qualitative corrections to the effect of plasmon enhanced
scattering described here. Thus, our main qualitative prediction of
the broad peak and the associated non-Lorentzian line shape of the
optical conductivity should remain valid independent of our neglect
of the many-body reconstruction of the Dirac spectrum, which is very
small at finite carrier densities by virtue of the fine structure
constant in graphene (particularly for graphene on substrates) being
not too large (of the order of unity or less)\ \cite{das_sarma_velocity_2013}.
Length scales involved in the plasmon-enhanced scattering are short
compared to the mean free path and therefore the disorder related
quantum interference effects give no contribution to the phenomenon
discussed in this paper.

The relatively large magnitude of the finite frequency peak in the
optical conductivity may enable an experimental confirmation of our
predictions. Moreover, the possibility to tune the position of the
plasmon induced peak in the optical conductivity of graphene, controlled
by the parameter $c=\sqrt{\alpha/(k_{F}d)}$, by tuning the Fermi
level in the system allows an unambiguous identification of the plasmon
enhanced dissipation effect in the optical response, which can otherwise
be masked by the interband transitions, scattering by phonons, and
other inelastic processes. Also, varying the location of impurities
$d$ allows additional tunability of the effect. This can be achieved
using hexagonal boron nitride spacers of different thicknesses $d$
between the graphene layer and $\textrm{Si\ensuremath{O_{2}}/Si}$
substrate with the latter containing the majority of charged impurities
in the vicinity of its surface. In particular, samples with relatively
wide hBN spacers $d\sim10\textrm{nm}$ would correspond to $c\ll1$
at densities $n\sim10^{12}\textrm{c\ensuremath{m^{-2}}}$ in which
case the plasmon induced peak appears at the relatively low frequency
$\hbar\omega\sim cE_{F}$ unobstructed by the interband transitions.
The plasmon enhanced dissipation discussed here is a generic disordered
Fermi-liquid phenomenon which nevertheless has not been unambiguously
identified experimentally so far, and graphene is an ideal system
to look for signatures of this phenomenon due to graphene's high electronic
quality and tunability of various system parameters, and the strong
magnitude of the plasmon-enhanced scattering in graphene.

With regard to the existing optical measurements the deviation of
the low frequency response from the Lorentzian shape has been reported
by some\ \cite{BasovFIR08,Basov12} but not all\ \cite{yan_infrared_2011,mak_optical_2012}
experiments indicating that the effect is sample dependent, which
is consistent with the mechanism (i.e. disorder plus plasmon) predicted
here. Redistribution of the spectral weight in the optical conductivity
and the resulting non-Lorentzian line shape of the Drude peak, if
present, have to be accounted for when extracting the Drude weight,
which has been recently used to analyze the spectrum renormalization
due to electron-electron interactions\ \cite{yan_infrared_2011,mak_optical_2012}.

The theoretical approach developed here is generic and can be straightforwardly
generalized to the case of the 2D Dirac fermions observed in other
materials such as topological insulators and 2D transition metal dichalcogenides.
The most important qualitative message of our theory is that the finite-frequency
dynamical conductivity appropriate for optical measurements in doped
graphene cannot be approximated to have a simple constant disorder
broadening given by the corresponding static transport scattering
rate because of the substantial interplay between plasmons and impurity
scattering. In particular, the finite-frequency scattering time appropriate
for the optical conductivity manifests very nontrivial frequency dependence,
which could produce unexpected qualitative phenomena such as the broad
finite-frequency peak in the conductivity with both the peak position
and the peak width being determined nontrivially by both impurity
scattering and Fermi energy. 
\begin{acknowledgments}
We thank Allan MacDonald and Euyheon Hwang for useful discussions.
This work is supported by US-ONR, LPS-CMTC, and JQI-NSF-PFC. 
\end{acknowledgments}

\begin{appendix}

\section{the band cut off in the low energy theory of graphene\label{sec:Appendix-I}}

Linearized Dirac spectrum contains an infinite number of states with
negative energies which cannot be realized in a condensed matter system.
Therefore a cut-off scheme is required to regularize the theory, which
may lead to a number of unphysical results\ \cite{mishchenko_effect_2007,mishchenko_minimal_2008,sheehy_quantum_2007}.
Also the spectral weight redistribution in such a model becomes ambiguous
as $f$-sum rule is cut-off dependent\ \cite{sabio_f-sum_2008,gusynin_sum_2007}.
Here we restrict our calculations to the physical quantities determined
by the low-energy states only. In particular, we neglect the effects
of many-body spectrum renormalization due to electron-electron interactions
keeping only the infrared divergent RPA diagrams. Therefore our results
are independent of the cutoff scheme and are expected to reproduce
the low-energy asymptotic of the realistic band structure calculation.

Formally, we introduce the hard cut-off which requires modification
of the field operators\ \cite{McDonaldEnhancedDrudeWeight} introduced
in Sec.\ \ref{sec:setup} of the main text,

\begin{eqnarray}
 & \hat{\Psi}_{\mathbf{k}}\rightarrow\theta(\Lambda-k)\hat{\Psi}_{\mathbf{k}},\label{eq:2ndQCutOff}\\
 & \rho(q)\rightarrow\mathrm{Tr}\sum_{\mathbf{k}}\theta(\Lambda-k)\theta(\Lambda-|\mathbf{k+q}|)\hat{\Psi}_{\mathbf{k}}^{\dagger}\hat{\Psi}_{\mathbf{k+q}}.
\end{eqnarray}
The sum over $\mathbf{k}$ is defined over the infinite range and
therefore allows the variable shift $\mathbf{k\rightarrow k+q}$.

We calculate the commutator $\left[H,\rho(q)\right]$

\begin{eqnarray}
 & \left[H_{0},\rho(q)\right]=-vq^{i}\mathrm{Tr}\sum_{\mathbf{k}}\hat{\Psi}_{\mathbf{k}}^{\dagger}\Sigma^{i}\hat{\Psi}_{\mathbf{k+q}},\\
 & \left[H_{imp},\rho(q)\right]=\mathrm{Tr}\sum_{\mathbf{k,q}}V_{\mathbf{q}}^{(i)}\left(\hat{\Psi}_{\mathbf{k-q}}^{\dagger}\hat{\Psi}_{\mathbf{k+q'}}\theta\left(\Lambda-k\right)\right.\nonumber \\
 & \left.-\hat{\Psi}_{\mathbf{k}}^{\dagger}\hat{\Psi}_{\mathbf{k+q+q'}}\theta\left(\Lambda-|\mathbf{k+q'}|\right)\right).
\end{eqnarray}

To include electron-electron interactions we use the mean-field approximation,
introducing,

\begin{eqnarray*}
 & d=\hat{\Psi}_{\mathbf{k+q}}^{\dagger}\hat{\Psi}_{\mathbf{k}}-\left\langle \hat{\Psi}_{\mathbf{k+q}}^{\dagger}\hat{\Psi}_{\mathbf{k}}\right\rangle ,
\end{eqnarray*}
and keeping only the first order in this parameter we arrive at,

\begin{eqnarray}
 & \left[H_{ee},\Psi_{\alpha\mathbf{k}}^{\dagger}\Psi_{\beta\mathbf{k+q'}}(t)\right]=\nonumber \\
 & =\sum_{q}V_{q}\rho(-q)\left[\Psi_{\alpha\mathbf{k-q}}^{\dagger}\Psi_{\beta\mathbf{k+q'}}\theta\left(\Lambda-k\right)\right.\nonumber \\
 & \left.-\Psi_{\alpha\mathbf{k}}^{\dagger}\Psi_{\beta\mathbf{k+q'+q}}\theta\left(\Lambda-|\mathbf{k+q'}|\right)\right].
\end{eqnarray}
The cut off gives rise to corrections to the theory presented in the
main text that vanish in the limit $\Lambda\rightarrow\infty$ and
therefore we dropped all cut off dependent terms in the main text.

\section{Full expression for the plasmon induced correction to the optical
conductivity \label{sub:Full-solution}}

Here we present the frequency-dependent contribution to the real part
of the optical conductivity proportional to the plasmon propagator.
This is obtained from Eq.\ (\ref{eq:Jpl}) using the solution of
the equation of motion in Eqs.\ (\ref{eq:g(11)}),\ (\ref{eq:Rho(11)})
and\ (\ref{eq:Resultg(01)}). Performing some tedious but straightforward
calculations we arrive at (here $\hbar\equiv1$),

\begin{eqnarray}
 & \Ree\sigma_{pl}=\frac{\eta^{2}e^{2}v^{2}\rho_{imp}}{2}\Imm\sum\frac{|V_{q}^{(i)}|^{2}V_{q}\left(AB+\frac{CD}{2}\right)}{\varepsilon^{2}(q,0)\varepsilon(q,\omega)},\label{eq:RPAFull}
\end{eqnarray}
where we introduced,

\begin{eqnarray*}
 & A\equiv-2\omega\sum_{\mathbf{k}}\mathbf{n_{k}\cdot n_{q}}\left[\frac{f_{\epsilon_{\mathbf{k+q}}}-f_{\epsilon_{\mathbf{k}}}}{\epsilon_{\mathbf{k+q}}-\epsilon_{\mathbf{k}}}\frac{\left(1+\mathbf{n_{k}\cdot n_{k+q}}\right)}{\omega^{2}-\left(\epsilon_{\mathbf{k+q}}-\epsilon_{\mathbf{k}}\right)^{2}}\right.\\
 & \left.+\frac{f_{\epsilon_{\mathbf{k+q}}}-f_{\epsilon_{\mathbf{k}}}}{\epsilon_{\mathbf{k}}+\epsilon_{\mathbf{k+q}}}\frac{\left(1-\mathbf{n_{k}\cdot n_{k+q}}\right)}{\left(\epsilon_{\mathbf{k}}+\epsilon_{\mathbf{k+q}}\right)^{2}-\omega^{2}}\right],
\end{eqnarray*}

\begin{eqnarray*}
 & B\equiv-2\sum_{\mathbf{k}}\mathbf{n_{k}\cdot n_{q}}\left[\frac{f_{\epsilon_{\mathbf{k+q}}}-f_{\epsilon_{\mathbf{k}}}}{\epsilon_{\mathbf{k}}+\epsilon_{\mathbf{k+q}}}\frac{\left(1-\mathbf{n_{k}\cdot n_{k+q}}\right)}{\omega\left(\left(\epsilon_{\mathbf{k}}+\epsilon_{\mathbf{k+q}}\right)^{2}-\omega^{2}\right)}\right.\\
 & \left.+\frac{f_{\epsilon_{\mathbf{k+q}}}-f_{\epsilon_{\mathbf{k}}}}{\epsilon_{\mathbf{k+q}}-\epsilon_{\mathbf{k}}}\left(\frac{\left(1+\mathbf{n_{k}\cdot n_{k+q}}\right)}{\omega^{2}-\left(\epsilon_{\mathbf{k}}-\epsilon_{\mathbf{k+q}}\right)^{2}}-\frac{\left(\epsilon_{\mathbf{k+q}}-\epsilon_{\mathbf{k}}\right)\left(1-\mathbf{n_{k}\cdot n_{k+q}}\right)}{2\omega\epsilon_{\mathbf{k}}\epsilon_{\mathbf{k+q}}}\right)\right],
\end{eqnarray*}

\begin{eqnarray*}
 & C=-2\sum_{k}\left(1-\left(\mathbf{n_{k}\cdot n_{q}}\right)^{2}\right)\frac{vq}{\epsilon_{\mathbf{k+q}}}\\
 & \times\left[\frac{f_{\epsilon_{\mathbf{k+q}}}-f_{\epsilon_{\mathbf{k}}}}{\epsilon_{\mathbf{k+q}}+\epsilon_{\mathbf{k}}}\frac{1}{\omega^{2}-\left(\epsilon_{\mathbf{k+q}}-\epsilon_{\mathbf{k}}\right)^{2}}+\frac{f_{\epsilon_{\mathbf{k+q}}}-f_{\epsilon_{\mathbf{k}}}}{\epsilon_{\mathbf{k+q}}-\epsilon_{\mathbf{k}}}\frac{1}{\omega^{2}-\left(\epsilon_{\mathbf{k+q}}+\epsilon_{\mathbf{k}}\right)^{2}}\right],
\end{eqnarray*}

\begin{eqnarray*}
 & D=-2\omega\sum_{\mathbf{k}}\left(1-\left(\mathbf{n_{k}\cdot n_{q}}\right)^{2}\right)\frac{vq}{\epsilon_{\mathbf{k+q}}}\\
 & \times\left[\frac{f_{\epsilon_{\mathbf{k+q}}}-f_{\epsilon_{\mathbf{k}}}}{\epsilon_{\mathbf{k+q}}-\epsilon_{\mathbf{k}}}\frac{\omega^{2}-2\epsilon_{\mathbf{k}}\left(\epsilon_{\mathbf{k+q}}-\epsilon_{\mathbf{k}}\right)}{\left(\omega^{2}-\left(\epsilon_{\mathbf{k+q}}-\epsilon_{\mathbf{k}}\right)^{2}\right)\left(\omega^{2}-4\epsilon_{\mathbf{k}}^{2}\right)}\right.\\
 & +\left.\frac{f_{\epsilon_{\mathbf{k+q}}}-f_{\epsilon_{\mathbf{k}}}}{\epsilon_{\mathbf{k}}+\epsilon_{\mathbf{k+q}}}\frac{\omega^{2}+2\epsilon_{\mathbf{k}}\left(\epsilon_{\mathbf{k+q}}+\epsilon_{\mathbf{k}}\right)}{\left(\omega^{2}-\left(\epsilon_{\mathbf{k+q}}+\epsilon_{\mathbf{k}}\right)^{2}\right)\left(\omega^{2}-4\epsilon_{\mathbf{k}}^{2}\right)}\right],
\end{eqnarray*}
where $\mathbf{n_{q}\equiv\mathbf{q}/}q$. 
\end{appendix}



%

\end{document}